\documentclass[12pt]{article}
\usepackage{amsmath}
\usepackage{amssymb}
\usepackage{amsfonts}

\newcommand{\field}[1]{\mathbb{#1}}
\newcommand{\R}{\field{R}}

\title{
Algebraic construction of associated functions of nondiagonalizable models with anharmonic oscillator complex interaction}
\author{I.\ MARQUETTE$^1$ and C.\ QUESNE$^{2}$\\ 
{\small $^1$ School of Mathematics and Physics, The University of Queensland,} \\ 
{\small Brisbane, QLD 4072, Australia}\\
{\small (e-mail: i.marquette@uq.edu.au)}\\ 
{\small $^2$ Physique Nucl\'eaireTh\'eorique et Physique Math\'ematique,  Universit\'e Libre de Bruxelles,} \\ 
{\small Campus de la Plaine CP229, Boulevard~du Triomphe, B-1050 Brussels, Belgium}\\
{\small (e-mail: Christiane.Quesne@ulb.be)}}
\date{ }
\begin{document}
\baselineskip=22pt plus 1pt minus 1pt
\maketitle

\begin{abstract} 
A shape invariant nonseparable and nondiagonalizable two-dimensional model with anharmonic complex interaction, first studied by Cannata, Ioffe, and Nishnianidze, is re-examined with the purpose of providing an algebraic construction of the associated functions to the excited-state wavefunctions, needed to complete the basis. The two operators $A^+$ and $A^-$, coming from the shape invariant supersymmetric approach, where $A^+$ acts as a raising operator while $A^-$ annihilates all wavefunctions, are completed by introducing a novel pair of operators $B^+$ and $B^-$, where $B^-$ acts as the missing lowering operator. It is then shown that building the associated functions as polynomials in $A^+$ and $B^+$ acting on the ground state provides a much more efficient approach than that used in the original paper. In particular, we have been able to extend the previous results obtained for the first two excited states of the quartic anharmonic oscillator either by considering the next three excited states or by adding a cubic or a sextic term to the Hamiltonian. 
\end{abstract}


\newpage

\section{Introduction}

Anharmonic oscillators play an important role in several branches of quantum physics as they describe for instance atomic vibrations of solids \cite{erba} and molecules \cite{carreira}, and also occur in field theory \cite{bender69}. To deal with such a problem, some approximation methods have often been used \cite{turbiner21a} or part of the spectrum has been derived for some specific values of the anharmonic term coefficients \cite{turbiner87, turbiner16}.\par
%
%
Some years ago, Cannata, Ioffe, and Nishnianidze \cite{cannata12} studied a class of quantum two-dimensional models with complex anharmonic oscillator interaction, described by pseudo-Hermitian Hamiltonians $H$, i.e., such that $\eta H \eta^{-1} = H^{\dagger}$ with $\eta$ a Hermitian invertible operator \cite{mosta02a, mosta10}. Such Hamiltonians generalize some previously studied one containing only harmonic terms \cite{cannata10}. Although those Hamiltonians turn out to be nonseparable, they are exactly solvable due to their property of shape invariance \cite{genden, cooper, junker, bagchi, cannata02, andrianov}.\par
%
%
The knowledge of their wavefunctions is, however, not enough to obtain a resolution of unity and a complete basis because the Hamiltonians are nondiagonalizable. It is therefore necessary to build some associated functions to the excited-state wavefunctions in order to complete the Jordan blocks and to get an extended biorthogonal basis, as well known for some one-dimensional pseudo-Hermitian Hamiltonians \cite{mosta02b, mosta03}. In \cite{cannata12}, the construction of associated functions was carried out in terms of the two complex variables $z$, $\bar{z}$ describing the system and was explicitly given only for the first two excited states in the case of quadratic plus quartic interactions. It indeed implied the rather difficult resolution of systems of coupled first-order differential equations, whose integration constants were found by eliminating possible nonphysical logarithmic solutions.\par
%
%
The purpose of the present work is to re-examine the associated function construction problem of \cite{cannata12} by adopting an algebraic approach based on the introduction of some novel ladder operators. Such a description will turn out to be easier because no logarithmic solution may appear and, as a consequence, it will enable us to extend the construction of associated functions to some higher excited states, as well as to more complicated interactions.\par
%
%
The paper is organized as follows. In Section~2, we review the model of \cite{cannata12} and construct a pair of new ladder operators. In Section~3, we discuss the problem of nondiagonalizability and introduce the method to be followed for algebraically constructing associated functions. In Section~4, such a construction is carried out in detail for the quartic anharmonic oscillator. It is extended to a quartic one with a cubic term and to a sextic anharmonic oscillator in Sections~5 and 6, respectively. Section~7 then contains the conclusion.\par
%
%
\section{Shape invariant model with anharmonic complex interaction}

\subsection{Shape invariant model and its exact solvability}

In \cite{cannata10}, it was observed that for $g = - (\omega_1^2 - \omega_2^2)/2$, the Schr\"odinger equation
\begin{equation}
  H \Psi({\bf x}) = E \Psi({\bf x}),  \label{eq:SE}
\end{equation}
corresponding to the two-dimensional model with complex oscillator Hamiltonian
\begin{equation}
  H = - \partial_1^2 - \partial_2^2 + \omega_1^2 x_1^2 + \omega_2^2 x_2^2 + 2 {\rm i} g x_1 x_2,
  \qquad \omega_1, \omega_2, g \in \R,
\end{equation}
or
\begin{equation}
  H = - 4\partial_z \partial_{\bar{z}} + 4a^2 z \bar{z} + 8ab \bar{z}^2, \qquad 2a = 
  \sqrt{\frac{1}{2}(\omega_1^2+ \omega_2^2)}, \qquad b = \frac{g}{8a},  \label{eq:HO}
\end{equation}
in terms of the complex variables $z = x_1+{\rm i}x_2$ and $\bar{z} = x_1-{\rm i}z_2$, cannot be separated into two differential equations by performing a linear complex transformation of variables because the Jacobian of the transformation then vanishes.\par
%
%
In \cite{cannata12}, such a remark was extended to the more general Hamiltonian
\begin{equation}
  H = - 4 \partial_z \partial_{\bar{z}} + 4a^2 z\bar{z} + 4a \bar{z} F'(\bar{z}),  \label{eq:AHO}
\end{equation}
where $F(\bar{z})$ is assumed to be some polynomial in $\bar{z}$ and $F'(\bar{z}) = \partial_{\bar{z}} F(\bar{z})$. As a matter of fact, for $F(\bar{z}) = b \bar{z}^2$, Hamiltonian (\ref{eq:AHO}) reduces to (\ref{eq:HO}). As its counterpart, the former satisfies pseudo-Hermiticity with $\eta$ chosen as $P_2$, where $P_2$ is the operator changing $x_2$ into $-x_2$.\par
%
%
Although nonseparable, the Schr\"odinger equation (\ref{eq:SE}) is however exactly solvable because with the operators
\begin{equation}
  A^{\pm} = \partial_z \mp a \bar{z},  \label{eq:A}
\end{equation}
Hamiltonian (\ref{eq:AHO}) satisfies the properties \cite{cannata12}
\begin{equation}
  H A^+ = A^+ (H+4a), \qquad A^- H = (H+4a) A^-,
\end{equation}
or
\begin{equation}
  [H, A^{\pm}] = \pm 4a A^{\pm},  \label{eq:H-A}
\end{equation}
characteristic of self-isospectral supersymmetry \cite{andrianov}. Hence, $H$ has an oscillator-like spectrum
\begin{equation}
  E_n = 4a (n+1), \qquad n=0, 1, 2, \ldots,  \label{eq:E}
\end{equation}
and its ground-state wavefunction is annihilated by $A^-$, while its excited-state ones are obtained from the latter by successive applications of $A^+$. The results read
\begin{equation}
  \Psi_{n,0}(z,\bar{z}) = c_{n,0} \bar{z}^n e^{-az\bar{z} - F(\bar{z})}, \qquad n=0, 1, 2, \ldots,
  \label{eq:wf}
\end{equation}
where $c_{n,0}$ is some normalization coefficient. Neither the spectrum (\ref{eq:E}), nor the corresponding wavefunctions (\ref{eq:wf}) depend on the detailed structure of the function $F(\bar{z})$.\par
%
%
The operator $A^+$ acts as a raising operator,
\begin{equation}
  A^+ \Psi_{n,0} = - 2a \frac{c_{n,0}}{c_{n+1,0}} \Psi_{n+1,0},
\end{equation}
but, in contrast with what would happen in the real case, $A^-$ is not a lowering operator since it annihilates not only the ground-state wavefunction, but also all the excited-state ones,
\begin{equation}
  A^- \Psi_{n,0} = 0, \qquad n=0, 1, 2, \ldots,
\end{equation}
and it has the unusual property of commuting with $A^+$,
\begin{equation}
  [A^-, A^+] = 0.
\end{equation}
\par
%
%
\subsection{Additional ladder operators}

In the case of the harmonic counterpart \cite{cannata10} of the anharmonic models considered here \cite{cannata12}, it has recently been shown \cite{marquette20a} that introducing an additional pair of ladder operators turns out to be very useful for exhibiting the hidden algebraic structure of the problem. Inspired by this result, let us consider here a novel set of operators
\begin{equation}
  B^{\pm} = \partial_{\bar{z}} \mp az \mp F'(\bar{z}),  \label{eq:B}
\end{equation}
where $B^-$ can provide us with the missing lowering operator. Its action on the set of wavefunctions $\Psi_{n,0}(z,\bar{z})$, defined in (\ref{eq:wf}), can indeed be easily shown to be given by
\begin{equation}
  B^- \Psi_{n,0} = n \frac{c_{n,0}}{c_{n-1,0}} \Psi_{n-1,0}.
\end{equation}
\par
%
%
This relation can be completed by the set of commutators
\begin{align}
  [H, B^{\pm}] &= \pm 4a B^{\pm} \pm 4 F^{\prime\prime}(\bar{z}) A^{\pm}, \label{eq:H-B}\\
  [B^-, B^+] &= -2 F^{\prime\prime}(\bar{z}),  \label{eq:B-B}
\end{align}
as well as
\begin{equation}
  [A^{\pm}, B^{\pm}] = 0, \qquad [A^{\pm}, B^{\mp}] = \pm 2a.
\end{equation}
Note that to express the right-hand side of (\ref{eq:H-B}) and (\ref{eq:B-B}) in terms of operators instead of the variable $\bar{z}$, we will need to give a precise definition of the polynomial $F(\bar{z})$, which will be done in the examples considered in Sections 4 to 6.\par
%
%
It is also possible to express the Hamiltonian in terms of the ladder operators in the following way
\begin{equation}
  H = 2(A^+ B^- + B^+ A^-) + 4a,
\end{equation}
which means that it admits an algebraic form independently of the anharmonicity terms.\par
%
%
\section{Nondiagonalizability and construction of associated functions}

\subsection{Nondiagonalizability and extended biorthogonal basis}

As well known, non-Hermitian Hamiltonians such as (\ref{eq:AHO}) need a suitable modification of the scalar product and resolution of identity \cite{mosta02a, mosta10, bender05, bender07}. A new scalar product can be defined as
\begin{equation}
  \langle\langle \Psi | \Phi\rangle\rangle = \int (P_2 T \Psi) \Phi d{\bf x},  \label{eq:sp}
\end{equation}
because $P_2 T$ is an appropriate antilinear operator commuting with $H$. Then, the pseudo-Hermitian Hamiltonian $H$ becomes Hermitian under the new scalar product (\ref{eq:sp}). Since the wavefunctions $\Psi_{n,0}(z,\bar{z})$ are simultaneously eigenfunctions of $P_2 T$ with unique eigenvalue $+1$, the new scalar product becomes an integral over the product of functions $\Psi \Phi$, instead of $\Psi^* \Phi$ as in quantum mechanics with real potentials.\par
%
%
As it was shown in \cite{cannata12}, the norms of the eigenstates (\ref{eq:wf}) are given by 
\begin{equation}
  \langle\langle \Psi_{n,0} | \Psi_{n,0}\rangle\rangle = \frac{\pi c_{n,0}^2}{2a} e^{-2F(0)} \delta_{n,0},
\end{equation}
which shows that all excited-state wavefunctions are self-orthogonal. This signals that one deals with a nondiagonalizable Hamiltonian \cite{mosta02b, mosta03}. As a consequence, some associated functions must be introduced to complete the basis and to get a resolution of identity. To $\Psi_{n,0}(z,\bar{z})$, $n \ge 1$, one has to add the functions $\Psi_{n,m}(z,\bar{z})$, $m=1$, 2, \ldots, $n$, defined by
\begin{equation}
  (H-E_n) \Psi_{n,m} = \Psi_{n,m-1}, \qquad m=1, 2, \ldots, n.  \label{eq:cond1}
\end{equation}
\par
%
%
Similarly, the partner eigenfunctions $\tilde{\Psi}_{n,0}(z,\bar{z})$, i.e., the eigenfunctions of $H^{\dagger}$, which are needed to complete a biorthogonal basis, are accompanied by their associated functions $\tilde{\Psi}_{n,m}(z,\bar{z})$, $m=1, 2, \ldots, n$, which can be taken as
\begin{equation}
  \tilde{\Psi}_{n,n-m} = \Psi^*_{n,m}, \qquad m=0, 1, \ldots, n.
\end{equation}
The scalar product in the extended biorthogonal basis is then
\begin{align}
  \langle \langle \Psi_{n,m} | \Psi_{n',m'} \rangle \rangle & = \langle \tilde{\Psi}_{n,m} | \Psi_{n',m'} \rangle
       = \int \Psi_{n,m} \Psi_{n',m'} d{\bf x} \nonumber \\
  & = \delta_{n,n'} \delta_{m,n-m'}, \qquad m=0, 1, \ldots, n, \quad m' = 0, 1, \ldots, n',  \label{eq:cond2}
\end{align}
with the corresponding decompositions
\begin{equation}
  I = \sum_{n=0}^{\infty} \sum_{m=0}^n | \Psi_{n,m} \rangle \rangle \langle \langle \Psi_{n,n-m} |
\end{equation}
and
\begin{equation}
  H = \sum_{n=0}^{\infty} \sum_{m=0}^n E_n | \Psi_{n,m} \rangle \rangle \langle \langle \Psi_{n,n-m} |
  + \sum_{n=0}^{\infty} \sum_{m=0}^{n-1} | \Psi_{n,m} \rangle \rangle \langle \langle \Psi_{n,n-m-1} |,
\end{equation}
showing that $H$ is block diagonal, each Jordan block having dimensionality $n+1$.\par
%
%
\subsection{Algebraic construction of associated functions}

Before going to the algebraic construction of associated functions for some specific anharmonic potentials, it is worth reviewing it for the oscillator one, corresponding to Hamiltonian (\ref{eq:HO}), because it will provide us with a clue to dealing with more general potentials.\par
%
%
{}For such a Hamiltonian and for the first excited-state wavefunction
\begin{equation}
  \Psi_{1,0} = - \frac{c_{1,0}}{2a} A^+ \Psi_0, \qquad \Psi_0 = e^{-az\bar{z} - b\bar{z}^2},  \label{eq:wf-1}
\end{equation}
one may look for an associated function in the form
\begin{equation}
  \Psi_{1,1} = (\mu A^+ + \nu B^+) \Psi_0,  \label{eq:assoc-1}
\end{equation}
where $\mu$ and $\nu$ are two yet unknown constants. The three constants $c_{1,0}$, $\mu$, and $\nu$ may be determined from (\ref{eq:cond1}) and (\ref{eq:cond2}). From (\ref{eq:cond1}) and the commutators (\ref{eq:H-A}) and (\ref{eq:H-B}), we get $\nu = - c_{1,0}/(16ab)$. From (\ref{eq:cond2}) for $m'=1$ and $m=0$, we obtain
\begin{align}
  1 &= \langle\langle \Psi_{1,0} | \Psi_{1,1}\rangle \rangle = \frac{c_{1,0}}{2a} \langle\Psi_0 | \eta A^-
       (\mu A^+ + \nu B^+)|\Psi_0\rangle = \frac{c_{1,0}}{2a} (-2a\nu) \langle\langle\Psi_0|\Psi_0\rangle\rangle   
       \nonumber \\
  &= \frac{c_{1,0}^2}{16ab}\langle\langle\Psi_0|\Psi_0\rangle\rangle,
\end{align}
leading to
\begin{equation}
  c_{1,0} = \left(\frac{16ab}{\langle\langle\Psi_0 | \Psi_0\rangle\rangle}\right)^{1/2}.  \label{eq:c-1,0}
\end{equation}
Finally, equation (\ref{eq:cond2}) with $m'=m=1$ yields
\begin{equation}
  0 = \langle\langle \Psi_{1,1} | \Psi_{1,1}\rangle\rangle = - \langle\Psi_0|\eta (\mu A^- + \nu B^-)(\mu A^+
  + \nu B^+)|\Psi_0\rangle = 4\nu(a\mu+b\nu) \langle\langle\Psi_0|\Psi_0\rangle\rangle.
\end{equation}
Hence $\mu = - (b/a) \nu$. The result for $\Psi_{1,1}$ therefore reads
\begin{equation}
  \Psi_{1,1} = \frac{c_{1,0}}{16a^2b} (bA^+ - aB^+) \Psi_0,
\end{equation}
with $c_{1,0}$ given in (\ref{eq:c-1,0}).\par
%
%
{}For higher values of $n$, the corresponding associated functions $\Psi_{n,m}$, $m=1$, 2, \ldots, $n$, can be found by taking appropriate linear combinations of $(A^+)^q (B^+)^{n-q} \Psi_0$, $q=0$, 1, \ldots, $n$. The results read
\begin{align}
  \Psi_{n,m} &= \frac{c_{n,0}}{(2b)^n (8a^2)^m m!} \sum_{q=n-m}^n (-1)^{m-q} \binom{m}{n-q} b^q
        a^{m-q} (A^+)^q (B^+)^{n-q} \Psi_0, \nonumber \\
  & \qquad  m=0, 1, \ldots, n,  \label{eq:assoc-ho-1}
\end{align}
where
\begin{equation}
  c_{n,0} = \left(\frac{(16ab)^n}{\langle\langle \Psi_0 | \Psi_0\rangle\rangle}\right)^{1/2}.
  \label{eq:assoc-ho-2}
\end{equation}
From the commutators of $H$, $A^{\pm}$, $B^{\pm}$, and simple properties of binomial coefficients, it is indeed a simple matter to show that Eqs.~(\ref{eq:cond1}) and (\ref{eq:cond2}) are satisfied by (\ref{eq:assoc-ho-1}) and (\ref{eq:assoc-ho-2}). Equation (\ref{eq:assoc-ho-1}) can also be rewritten in closed form as
\begin{equation}
  \Psi_{n,m} = \frac{c_{n,0}}{2^{n+3m} a^{n+m} b^m m!} (-A^+)^{n-m} (bA^+ - aB^+)^m \Psi_0.
\end{equation}
\par
%
%
\section{Quartic anharmonic oscillator}

For the quartic anharmonic oscillator, whose Hamiltonian is given by
\begin{equation}
  H = -4 \partial_z \partial_{\bar{z}} + 4a^2 z \bar{z} + 8ab \bar{z}^2 + 8a\omega \bar{z}^4,
  \label{eq:quartic}
\end{equation}
we get by comparison with Eq.(\ref{eq:AHO}), that the function $F(\bar{z})$ is given by $F(\bar{z}) = b \bar{z}^2 + \frac{\omega}{2} \bar{z}^4$. Hence, the corresponding additional ladder operators (\ref{eq:B}) become
\begin{equation}
  B^{\pm} = \partial_{\bar{z}} \mp az \mp 2\bar{z}(b + \omega \bar{z}^2)
\end{equation}
and the commutation relations (\ref{eq:H-B}) and (\ref{eq:B-B}) read
\begin{equation}
  [H, B^{\pm}] = \pm 4a B^{\pm} \pm 8b A^{\pm} \pm \frac{6\omega}{a^2} (A^+ - A^-)^2 A^{\pm}
\end{equation}
and
\begin{equation}
  [B^-, B^+] = - 4b - \frac{3\omega}{a^2} (A^+ - A^-)^2,
\end{equation}
where we have used (\ref{eq:A}) to express $\bar{z}$ in terms of $A^+$ and $A^-$.\par
%
%
The first excited-state wavefunction is given by an equation similar to (\ref{eq:wf-1}), except that $\Psi_0$ is now $\Psi_0 = \exp(-az\bar{z} - b\bar{z}^2 - \frac{\omega}{2}\bar{z}^4)$. For the corresponding associated function, we have to add some additional term to the right-hand side of (\ref{eq:assoc-1}). On assuming $\Psi_{1,1} = [\mu A^+ + \nu B^+ + \rho (A^+)^3] \Psi_0$, it is possible to satisfy both (\ref{eq:cond1}) and (\ref{eq:cond2}). By proceeding as in Section 3.2, we indeed get successively the conditions $\nu = - c_{1,0}/(16ab)$, $\rho = - (3\omega/4a^3)\nu$, $c_{1,0} = (16ab/\langle\langle \Psi_0 | \Psi_0 \rangle\rangle)^{1/2}$, and $\mu = - (b/a)\nu$. Hence, the final result reads
\begin{equation}
  \Psi_{1,1} = \frac{c_{1,0}}{16ab} \left[\frac{b}{a} A^+ - B^+ + \frac{3\omega}{4a^3} (A^+)^3\right]
  \Psi_0,
\end{equation}
with $c_{1,0}$ given by an equation similar to (\ref{eq:c-1,0}).\par
%
%
On proceeding in a similar way, we have found that up to $n=5$, the associated functions can be written as
\begin{align}
  \Psi_{n,m} &= \sum_{i=n-m+1}^{n-2} \alpha^{(n,m)}_{i, n-2-i} (A^+)^i (B^+)^{n-2-i} \Psi_0 \nonumber \\
  &\quad {}+ \sum_{k=0}^m \sum_{i=3k+n-m}^{n+2k} \alpha^{(n,m)}_{i,n+2k-i} (A^+)^i (B^+)^{n+2k-i}
       \Psi_0, \qquad m=1, 2, \ldots, n,
\end{align}
where the first summation on the right-hand side only exists for $m\ge 3$ and $\alpha^{(n,m)}_{i,j}$ are some nonvanishing coefficients, which can be written in the form
\begin{equation}
  \alpha^{n,m}_{i,j} = \frac{c_{n,0}}{2^{n+3} a^n b} \frac{N^{(n,m)}_{i,j}}{D^{(n,m)}_{i,j}}, \qquad
  c_{n,0} = \left(\frac{(16ab)^n}{\langle\langle \Psi_0 | \Psi_0\rangle\rangle}\right)^{1/2}.
  \label{eq:alpha}
\end{equation}
\par
%
%
By way of illustration, we list below the values of $\Psi_{n,m}$ for $n=2$, 3 and $m=1,2, \ldots, n$:
\begin{align}
  \Psi_{2,1} &= \frac{c_{2,0}}{32a^2b} \biggl\{\frac{3\omega-4b^2}{4ab} (A^+)^2 + A^+ B^+ 
      - \frac{3\omega}{4a^3} (A^+)^4\biggr\} \Psi_0, \\
  \Psi_{2,2} &= \frac{c_{2,0}}{32a^2b} \biggl\{- \frac{(3\omega-4b^2)(3\omega+4b^2)}{256a^2b^3} 
      (A^+)^2 - \frac{3\omega+4b^2}{32ab^2} A^+ B^+ + \frac{1}{16b} (B^+)^2 \nonumber \\
  &\quad {}+ \frac{3\omega(3\omega+4b^2)}{128a^4b^2} (A^+)^4 - \frac{3\omega}{32a^3b} (A^+)^3 B^+
      + \frac{9\omega^2}{256a^6b} (A^+)^6\biggr\} \Psi_0, \\
  \Psi_{3,1} &= \frac{c_{3,0}}{64a^3b} \biggl\{- \frac{3\omega-2b^2}{2ab} (A^+)^3 - (A^+)^2 B^+
      + \frac{3\omega}{4a^3} (A^+)^5\biggr\} \Psi_0, \\
  \Psi_{3,2} &= \frac{c_{3,0}}{64a^3b} \biggl\{\frac{3\omega(3\omega+2b^2)-4b^4}{64a^2b^3} (A^+)^3
      + \frac{1}{8a} (A^+)^2 B^+ - \frac{1}{16b} A^+ (B^+)^2 \nonumber \\
  &\quad {}- \frac{3\omega}{32a^4} (A^+)^5 + \frac{3\omega}{32a^3b} (A^+)^4 B^+ - \frac{9\omega^2}
      {256a^6b} (A^+)^7\biggr\} \Psi_0, \\
  \Psi_{3,3} &= \frac{c_{3,0}}{64a^3b} \biggl\{- \frac{\omega}{64ab^2} A^+ - \frac{81\omega^3 - 8b^6}
      {3072a^3b^5} (A^+)^3 - \frac{3\omega (3\omega+2b^2)+4b^4}{512a^2b^4} (A^+)^2 B^+
      \nonumber \\
  &\quad {}+ \frac{3\omega+2b^2}{256ab^3} A^+ (B^+)^2 - \frac{1}{384b^2} (B^+)^3 + 
      \frac{3\omega[3\omega(3\omega+2b^2)+4b^4]}{2048a^5b^4} (A^+)^5 \nonumber \\
  &\quad {}- \frac{3\omega(3\omega+2b^2)}{512a^4b^3} (A^+)^4 B^+ + \frac{3\omega}{512a^3b^2}
      (A^+)^3 (B^+)^2 + \frac{9\omega^2(3\omega+2b^2)}{4096a^7b^3} (A^+)^7 \nonumber \\
  &\quad{}- \frac{9\omega^2}{2048a^6b^2} (A^+)^6 B^+ + \frac{9\omega^3}{8192a^9b^2} (A^+)^9
      \biggr\} \Psi_0.
\end{align}
The expressions of $\Psi_{n,m}$ corresponding to $n=4$ and $n=5$ are given in Sections 8.1 and 8.2, respectively.\par
%
%
\section{Quartic anharmonic oscillator with a cubic term}

A first generalization of the quartic anharmonic oscillator, considered in \cite{cannata12} and in Section 4 above, consists in adding a cubic term in $\bar{z}$, so that $H$ becomes
\begin{equation}
  H = - 4\partial_z\partial_{\bar{z}} + 4a^2 z\bar{z} + 8ab \bar{z}^2 + 8a\epsilon \bar{z}^3 + 8a\omega
  \bar{z}^4,
\end{equation}
which corresponds to $F(\bar{z}) = b \bar{z}^2 + \frac{2\epsilon}{3} \bar{z}^3 + \frac{\omega}{2} \bar{z}^4$. The additional ladder operators (\ref{eq:B}) then read
\begin{equation}
  B^{\pm} = \partial_{\bar{z}} \mp az \mp 2\bar{z}(b + \epsilon \bar{z} + \omega \bar{z}^2)
\end{equation}
and lead to the commutation relations
\begin{equation}
  [H, B^{\pm}] = \pm 4a B^{\pm} \pm 8b A^{\pm} \pm \frac{8\epsilon}{a} (A^--A^+) A^{\pm} \pm
  \frac{6\omega}{a^2} (A^--A^+)^2 A^{\pm}
\end{equation}
and
\begin{equation}
  [B^-,B^+] = - 4b - \frac{4\epsilon}{a} (A^--A^+) - \frac{3\omega}{a^2} (A^--A^+)^2.
\end{equation}
\par
%
%
The wavefunctions of $H$ are given as before by successive applications of $A^+$ on $\Psi_0$, which now reads $\Psi_0 = \exp(-az\bar{z} - b \bar{z}^2 -\frac{2}{3}\epsilon \bar{z}^3 - \frac{\omega}{2} \bar{z}^4)$. The corresponding associated functions can be found by assuming some linear combinations of operators acting on $\Psi_0$ and finding the coefficients by solving (\ref{eq:cond1}) and (\ref{eq:cond2}). For $n=1$ and 2, the results are given by%
\begin{align}
  \Psi_{1,1} &= \frac{c_{1,0}}{16ab} \biggl\{\frac{b}{a} A^+ - B^+ - \frac{2\epsilon}{a^2} (A^+)^2
       + \frac{3\omega}{4a^3} (A^+)^3\biggr\} \Psi_0, \\
  \Psi_{2,1} &= \frac{c_{2,0}}{32a^2b} \biggl\{\frac{3\omega-4b^2}{4ab} (A^+)^2 + A^+ B^+ + 
       \frac{2\epsilon}{a^2} (A^+)^3 - \frac{3\omega}{4a^3} (A^+)^4\biggr\} \Psi_0, \\
  \Psi_{2,2} &= \frac{c_{2,0}}{32a^2b} \biggl\{- \frac{\epsilon}{4ab} A^+ - \frac{(3\omega-4b^2)(3\omega
       +4b^2)}{256a^2b^3} (A^+)^2 - \frac{3\omega+4b^2}{32ab^2} A^+ B^+ \nonumber \\
  &\quad {}+ \frac{1}{16b} (B^+)^2 - \frac{\epsilon(3\omega+4b^2)}{16a^3b^2} (A^+)^3 + \frac{\epsilon}
       {4a^2b} (A^+)^2 B^+ \nonumber \\
  &\quad {}+ \frac{3\omega(3\omega+4b^2)+32b\epsilon^2}{128a^4b^2} (A^+)^4 - \frac{3\omega}
       {32a^3b} (A^+)^3 B^+ \nonumber \\
  &\quad {}- \frac{3\omega\epsilon}{16a^5b} (A^+)^5 + \frac{9\omega^2}{256a^6b} (A^+)^6\biggr\}
       \Psi_0,
\end{align}
with $c_{n,0}$ expressed as in (\ref{eq:alpha}).\par
%
%
\section{Sextic anharmonic oscillator}

Another generalization of the quartic anharmonic oscillator previously considered consists in adding a sextic term in $\bar{z}$. The corresponding Hamiltonian reads
\begin{equation}
  H = - 4\partial_z\partial_{\bar{z}} + 4a^2 z\bar{z} + 8ab \bar{z}^2 + 8a\omega \bar{z}^4 + 8a\epsilon
  \bar{z}^6
\end{equation}
and is associated to the function $F(\bar{z}) = b \bar{z}^2 + \frac{\omega}{2} \bar{z}^4 + \frac{\epsilon}{3} \bar{z}^6$. The additional ladder operators (\ref{eq:B}) are now given by
\begin{equation}
  B^{\pm} = \partial_{\bar{z}} \mp az \mp 2\bar{z}(b + \omega \bar{z}^2 + \epsilon \bar{z}^4)
\end{equation}
and the corresponding commutation relations read
\begin{equation}
  [H, B^{\pm}] = \pm 4a B^{\pm} \pm 8b A^{\pm} \pm \frac{6\omega}{a^2}(A^+-A^-)^2 A^{\pm}
  \pm \frac{5\epsilon}{2a^4} (A^+-A^-)^4 A^{\pm}
\end{equation}
and
\begin{equation}
  [B^-, B^+] = - 4b - \frac{3\omega}{a^2} (A^+-A^-)^2 - \frac{5\epsilon}{4a^4} (A^+-A^-)^4.
\end{equation}
\par
%
%
{}For $n=1$ and 2, the corresponding associated functions can be written as
\begin{align}
  \Psi_{1,1} &= \frac{c_{1,0}}{16ab} \biggl\{\frac{b}{a} A^+ - B^+ + \frac{3\omega}{4a^3} (A^+)^3 +
      \frac{5\epsilon}{32a^5} (A^+)^5\biggr\} \Psi_0, \\
  \Psi_{2,1} &= \frac{c_{2,0}}{32a^2b} \biggl\{\frac{3\omega-4b^2}{4ab} (A^+)^2 + A^+ B^+
      - \frac{3\omega}{4a^3} (A^+)^4 - \frac{5\epsilon}{32a^5} (A^+)^6\biggr\} \Psi_0, \\
  \Psi_{2,2} &= \frac{c_{2,0}}{32a^2b} \biggl\{- \frac{(3\omega-4b^2)(3\omega+4b^2)}{256a^2b^3}
      (A^+)^2 - \frac{3\omega+4b^2}{32ab^2} A^+ B^+ + \frac{1}{16b} (B^+)^2 \\
  &\quad {}+ \frac{3\omega(3\omega+4b^2)-20b\epsilon}{128a^4b^2} (A^+)^4 - \frac{3\omega}{32a^3b}
      (A^+)^3 B^+ \nonumber \\
  &\quad {}+ \frac{36b\omega^2+5\epsilon(3\omega+4b^2)}{1024a^6b^2} (A^+)^6 - \frac{5\epsilon}
      {256a^5b} (A^+)^5 B^+ + \frac{15\omega\epsilon}{1024a^8b} (A^+)^8 \nonumber \\
  &\quad {} + \frac{25\epsilon^2}{16384a^{10}b} (A^+)^{10}\biggr\} \Psi_0,
\end{align}
where $\Psi_0 = \exp(-az\bar{z} - b \bar{z}^2 - \frac{\omega}{2} \bar{z}^4 - \frac{\epsilon}{3} \bar{z}^6)$ and $c_{n,0}$ has the same form as in (\ref{eq:alpha}).\par
%
%
\section{Conclusion}

In the present paper, we have re-examined a shape invariant nonseparable and nondiagonalizable two-dimensional model with anharmonic oscillator complex interaction, which was first studied in \cite{cannata12}, with the aim of providing an algebraic construction of the associated functions to the physical wavefunctions, needed to complete the basis and to get a resolution of identity.\par
%
%
{}For such a purpose, we have first made up for the lack of lowering operator coming from shape invariance by introducing a new operator $B^-$ possessing such a property. Together with the accompanying operator $B^+$, it has provided us with a pair of operators, completing the couple of operators $A^+$ and $A^-$ coming from the shape invariant supersymmetric approach of \cite{cannata12}.\par
%
%
We have then re-examined in detail the case of the quartic anharmonic oscillator, for which the associated functions to the first two excited states had been determined in \cite{cannata12} in terms of the variables $z$ and $\bar{z}$. We have shown that building them instead as polynomials in $A^+$ and $B^+$ acting on the ground state provides a much easier approach, which not only gives back the known results, but also allows to extend them up to the fifth excited state.\par
%
%
A similar procedure has been applied to two more general Hamiltonians, obtained by adding either a cubic or a sextic term to the quartic one. In both cases, the associated functions to the first two excited states have been determined in terms of the corresponding $A^+$ and $B^+$ operators.\par
%
%
In conclusion, we have established that our algebraic approach provides a very efficient method for building associated functions for arbitrarily high excited states and for arbitrarily complicated anharmonic interactions.\par
%
%
More generally, we would like to stress the interest of considering ladder operators in anharmonic oscillator problems. It is worth noting that other ladder operators have been constructed in terms of complex variables in the poly-analytic function theory, allowing applications to various models, such as modified Landau magnetic Hamiltonians \cite{turbiner21b}. Studying a possible description of ladder operators for anharmonic oscillator problems in such a framework would be an interesting topic for future investigation.\par
%
%
\section*{Acknowledgments}

I.M.\ was supported by Australian Research Council Fellowship FT180100099. C.Q.\ was supported by the Fonds de la Recherche Scientifique - FNRS under Grant Number 4.45.10.08.\par
%
%
\section{Appendix}

\subsection{Associated functions to the fourth excited state of the quartic anharmonic oscillator}

{}For $n=4$, the associated functions $\Psi_{n,m}$ corresponding to the quartic anharmonic oscillator (\ref{eq:quartic}) read
\begin{align}
  \Psi_{4,1} &= \frac{c_{4,0}}{128a^4b} \biggl\{\frac{9\omega-4b^2}{4ab} (A^+)^4 + (A^+)^3 B^+
       - \frac{3\omega}{4a^3} (A^+)^6\biggr\} \Psi_0, \\
  \Psi_{4,2} &= \frac{c_{4,0}}{128a^4b} \biggl\{- \frac{(3\omega+4b^2)(15\omega-4b^2)}{256a^2b^3}
       (A^+)^4 + \frac{3\omega-4b^2}{32ab^2} (A^+)^3 B^+ \nonumber \\
  &\quad {}+ \frac{1}{16b} (A^+)^2 (B^+)^2 - \frac{3\omega(3\omega-4b^2)}{128a^4b^2} (A^+)^6 
       - \frac{3\omega}{32a^3b} (A^+)^5 B^+ \nonumber  \\
  &\quad {}+ \frac{9\omega^2}{256a^6b} (A^+)^8\biggr\} \Psi_0, \\
  \Psi_{4,3} &= \frac{c_{4,0}}{128a^4b} \biggl\{\frac{\omega}{64ab^2} (A^+)^2 + \frac{1215\omega^3
       +324b^2\omega^2 + 144b^4\omega-64b^6}{24576a^3b^5} (A^+)^4 \nonumber \\
  &\quad {}- \frac{(3\omega+4b^2)(3\omega-4b^2)}{2048a^2b^4} (A^+)^3 B^+ - \frac{3\omega+4b^2}
       {512ab^3} (A^+)^2 (B^+)^2 + \frac{1}{384b^2} A^+ (B^+)^3 \nonumber \\
  &\quad {}+ \frac{3\omega(3\omega+4b^2)(3\omega-4b^2)}{8192a^5b^4} (A^+)^6 + \frac{3\omega
       (3\omega+4b^2)}{1024a^4b^3} (A^+)^5 B^+ \nonumber \\
  &\quad {}- \frac{3\omega}{512a^3b^2} (A^+)^4 (B^+)^2 - \frac{9\omega^2(3\omega+4b^2)}{8192
       a^7b^3} (A^+)^8 + \frac{9\omega^2}{2048a^6b^2} (A^+)^7 B^+ \nonumber \\
  &\quad {}- \frac{9\omega^3}{8192a^9b^2} (A^+)^{10} \biggr\} \Psi_0, \\
  \Psi_{4,4} &= \frac{c_{4,0}}{128a^4b} \biggl\{- \frac{\omega(9\omega+4b^2)}{2048a^2b^4} (A^+)^2
       + \frac{\omega}{512ab^3} A^+ B^+ \nonumber \\
  &\quad {}- \frac{49815\omega^4+5184b^2\omega^3+2208b^4\omega^2-256b^8}{3145728a^4b^7}
       (A^+)^4 \nonumber \\
  &\quad {}- \frac{567\omega^3+540b^2\omega^2+144b^4\omega+64b^6}{196608a^3b^6} (A^+)^3
       B^+ \nonumber \\
  &\quad {}+ \frac{99\omega^2+48b^2\omega+16b^4}{32768a^2b^5} (A^+)^2 (B^+)^2
       - \frac{9\omega+4b^2}{12288ab^4} A^+ (B^+)^3 \nonumber \\
  &\quad {}+ \frac{1}{12288b^3} (B^+)^4 + \frac{\omega(567\omega^3+540b^2\omega^2+144b^4\omega
       +64b^6)}{262144a^6b^6} (A^+)^6 \nonumber \\
  &\quad {}- \frac{3\omega(99\omega^2+48b^2\omega+16b^4)}{65536a^5b^5} (A^+)^5 B^+
       + \frac{3\omega(9\omega+4b^2)}{16384a^4b^4} (A^+)^4 (B^+)^2 \nonumber \\
  &\quad {}- \frac{\omega}{4096a^3b^3} (A^+)^3 (B^+)^3 + \frac{9\omega^2(99\omega^2+48b^2\omega
       +16b^4)}{524288a^8b^5} (A^+)^8 \nonumber\\
  &\quad {}- \frac{9\omega^2(9\omega+4b^2)}{65536a^7b^4} (A^+)^7 B^+ + \frac{9\omega^2}{32768
       a^6b^3} (A^+)^6 (B^+)^2 \nonumber \\
  &\quad {}+ \frac{9\omega^3(9\omega+4b^2)}{262144a^{10}b^4} (A^+)^{10} - \frac{9\omega^3}
       {65536a^9b^3} (A^+)^9 B^+ + \frac{27\omega^4}{1048576a^{12}b^3} (A^+)^{12}\biggr\} \Psi_0.       
\end{align}
\par
%
%
\subsection{Associated functions to the fifth excited state of the quartic anharmonic oscillator}

{}For $n=5$, the associated functions $\Psi_{n,m}$ corresponding to the quartic anharmonic oscillator (\ref{eq:quartic}) read
\begin{align}
  \Psi_{5,1} &= \frac{c_{5,0}}{256a^5b} \biggl\{- \frac{3\omega-b^2}{ab} (A^+)^5 - (A^+)^4 B^+
       + \frac{3\omega}{4a^3} (A^+)^7\biggr\} \Psi_0, \\
  \Psi_{5,2} &= \frac{c_{5,0}}{256a^5b} \biggl\{\frac{9\omega^2+18b^2\omega-4b^4}{64a^2b^3} (A^+)^5
       - \frac{3\omega-2b^2}{16ab^2} (A^+)^4 B^+ - \frac{1}{16b} (A^+)^3 (B^+)^2 \nonumber \\
  &\quad {}+ \frac{3\omega(3\omega-2b)}{64a^4b^2} (A^+)^7 + \frac{3\omega}{32a^3b} (A^+)^6 B^+
       - \frac{9\omega^2}{256a^6b} (A^+)^9\biggr\} \Psi_0, \\
  \Psi_{5,3} &= \frac{c_{5,0}}{256a^5b} \biggl\{- \frac{\omega}{64ab^2} (A^+)^3 - \frac{81\omega^3
      +27b^2\omega^2+18b^4\omega-4b^6}{1536a^3b^5} (A^+)^5 \nonumber \\
  &\quad {}+ \frac{9\omega^2+6b^2\omega-4b^4}{512a^2b^4} (A^+)^4 B^+ + \frac{1}{128ab} (A^+)^3
      (B^+)^2 - \frac{1}{384b^2} (A^+)^2 (B^+)^3 \nonumber \\
  &\quad {}- \frac{3\omega(9\omega^2+6b^2\omega-4b^4)}{2048a^5b^4} (A^+)^7 - \frac{3\omega}
      {256a^4b} (A^+)^6 B^+ + \frac{3\omega}{512a^3b^2} (A^+)^5 (B^+)^2 \nonumber \\
  &\quad {}+ \frac{9\omega^2}{2048a^7b} (A^+)^9 - \frac{9\omega^2}{2048a^6b^2} (A^+)^8 B^+
      + \frac{9\omega^3}{8192a^9b^2} (A^+)^{11} \biggr\} \Psi_0, \\
  \Psi_{5,4} &= \frac{c_{5,0}}{256a^5b} \biggl\{\frac{\omega(3\omega+2b^2)}{1024a^2b^4} (A^+)^3
      - \frac{\omega}{512ab^3} (A^+)^2 B^+ \nonumber \\
  &\quad {}+ \frac{4131\omega^4+972b^2\omega^3+1368b^4\omega^2+192b^6\omega-16b^8}
      {196608a^4b^7} (A^+)^5 \nonumber \\
  &\quad {}- \frac{81\omega^3-8b^6}{24576a^3b^6} (A^+)^4 B^+ - \frac{9\omega^2+6b^2\omega+4b^4}
      {8192a^2b^5} (A^+)^3 (B^+)^2 \nonumber \\
  &\quad {}+ \frac{3\omega+2b^2}{6144ab^4} (A^+)^2 (B^+)^3 - \frac{1}{12288b^3} A^+ (B^+)^4
      + \frac{\omega(81\omega^3-8b^6)}{32768a^6b^6} (A^+)^7 \nonumber \\
  &\quad {}+ \frac{3\omega(9\omega^2+6b^2\omega+4b^4)}{16384a^5b^5} (A^+)^6 B^+ - \frac{3\omega
      (3\omega+2b^2)}{8192a^4b^4} (A^+)^5 (B^+)^2 \nonumber \\
  &\quad {}+ \frac{\omega}{4096a^3b^3} (A^+)^4 (B^+)^3 - \frac{9\omega^2(9\omega^2+6b^2\omega
      +4b^4)}{131072a^8b^5} (A^+)^9 \nonumber \\
  &\quad {}+ \frac{9\omega^2(3\omega+2b^2)}{32768a^7b^4} (A^+)^8 B^+ - \frac{9\omega^2}
      {32768a^6b^3} (A^+)^7 (B^+)^2 \nonumber \\
  &\quad {}- \frac{9\omega^3(3\omega+2b^2)}{131072a^{10}b^4} (A^+)^{11} + \frac{9\omega^3}
      {65536a^9b^3} (A^+)^{10} B^+ - \frac{27\omega^4}{1048576a^{12}b^3} (A^+)^{13} \biggr\} \Psi_0, 
\end{align}
\begin{align}
  \Psi_{5,5} &= \frac{c_{5,0}}{256a^5b} \biggl\{- \frac{\omega(45\omega^2+18b^2\omega+4b^4)}
      {32768a^3b^6} (A^+)^3 + \frac{\omega(3\omega+b^2)}{4096a^2b^5} (A^+)^2 B^+ \nonumber \\
  &\quad {} - \frac{\omega}{8192ab^4} A^+ (B^+)^2 \nonumber \\  
  &\quad {}- \frac{69255\omega^5+10935b^2\omega^4-12420b^4\omega^3+4680b^6\omega^2
      +720b^8\omega-16b^{10}}{7864320a^5b^9} (A^+)^5 \nonumber \\
  &\quad {}+ \frac{243\omega^4-972b^2\omega^3+936b^4\omega^2+96b^6\omega-16b^8}{1572864
      a^4b^4} (A^+)^4 B^+ \nonumber \\
  &\quad {}+ \frac{162\omega^3+81b^2\omega^2+18b^4\omega+4b^6}{196608a^3b^7} (A^+)^3 (B^+)^2
      \nonumber \\
  &\quad {}- \frac{45\omega^2+18b^2\omega+4b^4}{196608a^2b^6} (A^+)^2 (B^+)^3 + \frac{3\omega
      +b^2}{98304ab^5} A^+ (B^+)^4 - \frac{1}{491520b^4} (B^+)^5 \nonumber \\
  &\quad {}- \frac{\omega(243\omega^4-972b^2\omega^3+792b^4\omega^2+96b^6\omega-16b^8}
      {2097152a^7b^8} (A^+)^7 \nonumber \\
  &\quad {}- \frac{\omega(162\omega^3+81b^2\omega^2+18b^4\omega+4b^6}{131072a^6b^7} (A^+)^6
      B^+ \nonumber \\
  &\quad {}+ \frac{3\omega(45\omega^2+18b^2\omega+4b^4)}{262144a^5b^6} (A^+)^5 (B^+)^2
      - \frac{\omega(3\omega+b^2)}{32768a^4b^5} (A^+)^4 (B^+)^3 \nonumber \\
  &\quad {}+ \frac{\omega}{131072a^3b^4} (A^+)^3 (B^+)^4  + \frac{3\omega^2(162\omega^3 
      +81b^2\omega^2+18b^4\omega+4b^6)}{1048576a^9b^7} (A^+)^9 \nonumber \\
  &\quad {}- \frac{9\omega^2(45\omega^2+18b^2\omega+8b^4)}{1048576a^8b^6} (A^+)^8 B^+
      + \frac{9\omega^2(3\omega+b^2)}{262144a^7b^5} (A^+)^7 (B^+)^2 \nonumber \\
  &\quad {}- \frac{3\omega^2}{262144a^6b^4} (A^+)^6 (B^+)^3 + \frac{9\omega^3(45\omega^2
      +18b^2\omega+4b^4)}{4194304a^{11}b^6} (A^+)^{11} \nonumber \\
  &\quad {}- \frac{9\omega^3(3\omega+b^2)}{524288a^{10}b^5} (A^+)^{10} B^+ + \frac{9\omega^3}
      {1048576a^9b^4} (A^+)^9 (B^+)^2 \nonumber \\
  &\quad {}+ \frac{27\omega^4(3\omega+b^2)}{8388608a^{13}b^5} (A^+)^{13} - \frac{27\omega^4}
      {8388608a^{12}b^4} (A^+)^{12} B^+ \nonumber \\
  &\quad {}+ \frac{81\omega^5}{167772160a^{15}b^4} (A^+)^{15}\biggr\} \Psi_0.
\end{align}
\par
%
%


\newpage

\begin{thebibliography}{99}

\bibitem{erba}
A.~Erba, J.~Maul, M.~Ferrabone, P.~Carbonni\`ere, M.~R\'erat and R.~Dovesi: 
Anharmonic vibrational states of solids from DFT calculations. Part I. Description of the potential energy surface, 
{\em J.~Chem.~Theory Comput.\/} {\bf 15}, 3755 (2019).

\bibitem{carreira}
L.~A.~Carreira, I.~M.~Mills and W.~B.~Person:
Two-dimensional anharmonic oscillator. Application to 2,5-dihydrofuran,
{\em J.~Chem.~Phys.\/} {\bf 56}, 1444 (1972).

\bibitem{bender69}
C.~M.~Bender and T.~T.~Mu:
Anharmonic oscillator,
{\em Phys.~Rev.\/} {\bf 184}, 1231 (1969). 

\bibitem{turbiner21a}
A.~V.~Turbiner and J.~C.~del Valle:
Anharmonic oscillator: a solution,
{\em J.~Phys.~A\/} {\bf 54}, 295204 (2021).

\bibitem{turbiner87}
A.~V.~Turbiner and A.~G.~Ushveridze:
Spectral singularities and quasi-exactly solvable quantal problem,
{\em Phys~ Lett.~A\/} {\bf 126}, 181 (1987). 

\bibitem{turbiner16}
A.~V.~Turbiner:
One-dimensional quasi-exactly solvable Schr\"odinger equations,
{\em Phys.~Rep.\/} {\bf 642}, 1 (2016).

\bibitem{cannata12}
F.~Cannata, M.~V.~Ioffe and D.~N.~Nishnianidze:
Equidistance of the complex two-dimensional anharmonic oscillator spectrum: the exact solution,
{\em J.~Phys.~A\/} {\bf 45}, 295303 (2012).

\bibitem{mosta02a}
A.~Mostafazadeh:
Pseudo-Hermiticity versus PT symmetry: The necessary condition for the reality of the spectrum of a non-Hermitian Hamiltonian,
{\em J.~Math.~Phys.\/} {\bf 43}, 205 (2002).

\bibitem{mosta10}
A.~Mostafazadeh:
Pseudo-Hermitian representation of quantum mechanics,
{\em Int.~J.~Geom.~Meth.~Mod.~Phys.\/} {\bf 7}, 1191 (2010).

\bibitem{cannata10}
F.~Cannata, M.~V.~Ioffe and D.~N.~Nishnianidze:
Exactly solvable nonseparable and nondiagonalizable two-dimensional model with quadratic complex interaction,
{\em J.~Math.~Phys.\/} {\bf 51}, 022108 (2010).

\bibitem{genden}
L.~E.~Gendenshtein:
Derivation of exact spectra of the Schr\"odinger equation by means of supersymmetry,
{\em JETP~Lett.\/} {\bf 38}, 356 (1983).

\bibitem{cooper}
F.~Cooper, A.~Khare and U.~Sukhatme:
Supersymmetry and quantum mechanics,
{\em Phys.~Rep.\/} {\bf25}, 268 (1995).

\bibitem{junker}
G.~Junker:
{\em Supersymmetric Methods in Quantum and Statistical Physics},
Springer-Verlag, Berlin, 1996.

\bibitem{bagchi}
B.~K.~Bagchi:
{\em Supersymmetry in Quantum and Classical Physics},
Chapman \& Hall/CRC, Boca Raton, FL, 2001.

\bibitem{cannata02}
F.~Cannata, M.~V.~Ioffe and D.~N.~Nishnianidze:
New methods for the two-dimensional Schr\"odinger equation: SUSY-separation of variables and shape invariance,
{\em J.~Phys.~A\/} {\bf35}, 1389 (2002).

\bibitem{andrianov}
A.~A.~Andrianov, F.~Cannata, M.~V.~Ioffe and D.~N.~Nishnianidze:
Systems with higher-order shape invariance: spectral and algebraic properties,
{\em Phys.~Lett~A\/} {\bf 266}, 341 (2000).

\bibitem{mosta02b}
A.~Mostafazadeh:
Pseudo-Hermiticity for a class of nondiagonalizable Hamiltonians,
{\em J.~Math.~Phys.\/} {\bf 43}, 6343 (2002).

\bibitem{mosta03}
A.~Mostafazadeh:
Erratum: Pseudo-Hermiticity for a class of nondiagonalizable Hamiltonians [J.~Math.~Phys.~43, 6343 (2002)],
{\em J.~Math.~Phys.\/} {\bf 44}, 943 (2003).

\bibitem{marquette20a}
I.~Marquette and C.~Quesne:
Ladder operators and hidden algebras for shape invariant nonseparable and nondiagonalizable models with quadratic complex interaction. I. Two-dimensional model, 
{\em SIGMA\/} {\bf 18}, 004 (2022).

\bibitem{bender05}
C.~M.~Bender:
Introduction to $ \cal PT$-symmetric quantum theory,
{\em Contemp.~Phys.\/} {\bf 46}, 277 (2005).

\bibitem{bender07}
C.~M.~Bender:
Making sense of non-Hermitian Hamiltonians,
{\em Rep.~Prog.~Phys.\/} {\bf70}, 947 (2007).

\bibitem{turbiner21b}
A.~V.~Turbiner and N.~Vasilevski:
Poly-analytic functions and representation theory,
{\em Complex~Analysis~and~Operator~Theory\/} {\bf 15}, 110 (2021). 

\end{thebibliography}
\end{document}